\documentclass[preprint,authoryear,12pt]{elsarticle}
 \usepackage{amssymb}
 \usepackage{amsthm}
 \usepackage{natbib}
 \usepackage{amsmath}
 \usepackage{amsmath,amssymb,amsthm,graphicx,caption}
 \usepackage{amsfonts}
 \usepackage{mathrsfs}
 \usepackage{verbatim}
 \usepackage{bm}
 \usepackage[usenames, dvipsnames]{color}
 \usepackage[capposition=top]{floatrow}
 \usepackage[section]{placeins}
 \usepackage[utf8]{inputenc}
 \usepackage{rotating}
 \usepackage[figurename=Table]{caption}
 \usepackage[colorlinks=true,linkcolor=blue]{hyperref}%
 \usepackage{lscape}

  \newcommand{\Lower}[1]{\smash{\lower 1.5ex \hbox{#1}}}

  \def\vep{\varepsilon}
  
  \def\ba{{\mathbf{a}}}
  \def\R{\mathbb{R}}

  \def\bW{\mathbf{W}}

  \def\pd{\frac{\partial Q_n(\theta(\lambda,h))}{\partial\theta}}
  \def\pdd{\frac{\partial^2 Q_n(\theta(\lambda,h))}{\partial\theta\partial\theta^T}}

  \newenvironment{eqarray}{\arraycolsep 0.14em\begin{eqnarray}}{\end{eqnarray}}
  \newenvironment{eqarray*}{\arraycolsep 0.14em\begin{eqnarray*}}{\end{eqnarray*}}
  \newtheorem{thm}{Theorem}

 \journal{Submitted to TBA}

\begin{document}

\begin{frontmatter}

\title{{\bf\large SIMEX Estimation in Parametric Modal Regression with Measurement Error}}
\author{Jianhong Shi, Yujing Zhang, Ping Yu}
\address{School of Mathematics and Computer Sciences, Shanxi Normal University, Linfen, China 041000}
\author{Weixing Song\corref{cor}}
\ead{weixing@ksu.edu}
\cortext[cor]{Corresponding author}
\address{Department of Statistics, Kansas State University, Manhattan, KS 66506}

\begin{abstract}
  For a class of parametric modal regression models with measurement error, a simulation extrapolation estimation procedure is proposed in this paper for estimating the modal regression coefficients. Large sample properties of the proposed estimation procedure, including the consistency and asymptotic normality, are thoroughly investigated. Simulation studies are conducted to evaluate its robustness to potential outliers and the effectiveness in reducing the bias caused by the measurement error.
\end{abstract}

\end{frontmatter}

%% \linenumbers

%----------------------------------------------------------------------

 \section{Introduction}

 Modal or mode regression, together with the mean and quantile regression, provides data analysts a suite of inference tools to describe the data structures and to model the relationships among variables. Comparing to the well developed mean and quantile regression techniques, the modal regression is still expanding its territory in applications and theory. The modal estimation idea germinated over half century ago in \cite{Parzen1962On} and \cite{Chernoff1964Estimation} on estimating the mode of a probability density function. Later, similar ideas were extended to regression setups. For example, \cite{Sager1982Maximum} discussed the maximum likelihood estimation in isotonic mode regression. To our best knowledge, it is \cite{Lee1989Mode} who considered the linear modal regression by minimizing a proper risk function, and as a further development, \cite{Lee1993Quadratic} reformulated the estimation procedure using the rectangular kernel and the Epanechnikov kernel. However, in both works, the bandwidths are fixed, the consistency and the asymptotic normality are achieved by requiring the density function of the response variable given the predictors to be symmetric about the mode, at least up to plus and minus the bandwidth. The kernel idea developed in \cite{Lee1989Mode, Lee1993Quadratic} was eventually refined in \cite{Kemp2012Regression} where the modal regression estimate was formally defined as the maximizer of the kernel density estimate of the regression error evaluated at the origin. Independent of \cite{Kemp2012Regression}'s work, \cite{Yao2014A} also discussed the same estimation procedure. In addition to the similar large sample results, \cite{Yao2014A} developed the breakdown point theory of the proposed estimator and provided a data-driven bandwidth selector. Recently, \cite{Khardani2017Non} discussed the modal regression in non-linear setups, weak convergence and asymptotic normality of the modal regression coefficient estimators are investigated.

 The above mentioned literature assume that all variables in the regression models are observable. However, in real applications, some variables cannot be measured precisely due to various reasons. Such examples can be easily found in econometrics, biology, nutrition and toxicology studies, see \cite{Carroll2006measurement} for more examples. Extensive research has been conducted for the quantile and other traditional robust statistical inference procedures in the measurement error setup, only recently have we witnessed increasing interest in modal regression models when the covariates are contaminated with measurement errors. \cite{Li2019} considers the linear mode regression in the presence of measurement errors and proposes two estimation methods, the corrected score and the corrected kernel estimators. The correct score estimator is an application of \cite{novick2002}'s estimation procedure by assuming that the measurement error has a normal distribution and the estimating function is entire with respect to the predictors. In fact, the corrected score estimator proposed in \cite{novick2002} and \cite{Li2019} is a variant of SIMEX estimation procedure. The corrected kernel estimator is indeed the deconvolution kernel estimator. Realizing that the modal residual in linear regression after plugging in the surrogate variable is a convolution of the regression error and the measurement error, a deconvolution kernel density estimator for the modal regression residual is constructed, and the modal regression coefficients are then estimated by maximizing the deconvolution kernel density function. Large sample properties are derived when the measurement error follows ordinary and super smooth distributions. In nonparametric setup, \cite{Zhou2016nonparametric} discussed the modal regression in the presence of measurement error by considering a mixture of classical and deconvolution kernel estimate for the joint distribution of the response and predictors.

 In this paper, we will focus on the estimation in a class of parametric modal regression when the covarites are observed with measurement errors. To reduce the potential biases introduced by the measurement error, we attempt to apply the classical SIMEX procedure to estimate the regression coefficients. The commonly used corrected score method should be considered if the corrected score function can be explicitly obtained. However, in real applications, the correct score function is often very hard to derive and in this case, SIMEX is an ideal alternative. In particular, the score functions based on the measurement error free data are often well established, and recent decades have seen a fast development in computing capability, these make using SIMEX an efficient way to estimate unknown parameters in most statistical models involving the measurement errors.

 The paper is organized as follows. The parametric modal regression model with measurement error and the SIMEX estimation procedure will be introduced in Section 2; large sample properties of the proposed estimator will be discussed in Section 3. Finally, simulation studies are conducted in Section 4 to evaluate the finite sample performance of the proposed SIMEX estimator. All the proofs of the main results are deferred to Appendix.

 Throughout this paper, the following notations will be used. For a generic function $g(x;\theta)$, where $x$ is the argument and $\theta$ is a parameter, possibly multidimensional, the first two derivatives of $g$ with respect to $x$ are denoted by $f'(x;\theta)$ and $f''(x;\theta)$, and the first two derivatives of $g$ with respect to $\theta$ are denoted by $\dot g(x;\theta)$ or $\ddot g(x;\theta)$, respectively. For any vector or matrix $A$, we use $A^{\otimes 2}$ to denote $AA^T$, where $A^T$ is the transpose of $A$. For the sake of simplicity, the multiple integration will be denoted by a single integration sign, and for a $p$-dimensional vector $u$, $du=du_1\cdots du_k$.

 \section{Parametric EV Modal Regression Using SIMEX}\label{sec2}

 To be specific, the parametric modal regression model with measurement error to be discussed in this paper takes the form
   \begin{equation}\label{eq2.1}
     Y=m(X,\theta)+\vep,\quad W=X+U,
   \end{equation}
 where $Y$ is a $1$-dimensional response variable, the true predictor $X$, being a $p$-dimensional real random vector, cannot be observed directly. What we have are the observations from $W$, which is related to $X$ through the additive relationship $W=X+U$ with $U$ being the measurement error, and independent of $X$ and $\vep$. $\theta$ is a $q$-dimensional unknown vector of parameters to be estimated. We further assume that the measurement error $U\sim N_p(0,\Sigma_u)$, and $\Sigma_u$ is a known positive definite matrix.

 The key assumption in modal regression is that the marginal density function $g(\vep)$ of $\vep$ has a unique mode at $0$. When both $(Y,X)$ are available, then $g(0)$, the density function $g(\vep)$ at $0$, can be estimated by
   \begin{equation}\label{eq2.1.1}
     \hat g_n(0,\theta)=\frac{1}{nh}\sum_{i=1}^n K\left(\frac{Y_i-m(X_i,\theta)}{h}\right).
   \end{equation}
 The modal estimate of $\theta$ is defined as the maximizer of $\hat g_n(0,\theta)$. As for the reasons why this procedure produce a reasonable estimate for $\theta$, see \cite{Yao2014A}. It is noted that $\hat g_n(0,\theta)$ defined above is not the kernel estimate of conditional density function of $\vep$ given $X$, but rather the kernel estimate of the marginal density function of $\vep$,  evaluated at $0$.  In the measurement error setup, due to the unavailability of observations on $X$, one can not maximize $\hat g_n(0,\theta)$ to get the modal estimate of $\theta$. The naive procedure by simply replacing $X_i$'s with $W_i$'s in the expression of $\hat g_n(0,\theta)$ has been proven to be an undesirable action in that the resulting estimate are often biased and as a consequence, the statistical inferences based on the naive estimate are often invalid.

 One may consider a corrected score type of methods to avoid the potential bias induced by the measurement error, that is, find a proper function of $(Y,Z)$ and $\theta$, say $S(Y,Z,\theta)$ such that $Eh^{-1}K((Y-m(X,\theta))/h)=ES(Y,Z,\theta)$. However, unless in some very special cases, such as the measurement error has a multivariate Laplace distribution or the modal regression function $m$ has some particular forms, constructing such functions often poses great challenges, if not infeasible. In the following, we shall design a SIMEX estimation procedure to estimate the modal regression parameter $\theta$. The significance of SIMEX is that one can simply rely on computer and a standard estimation procedure based on $(Y,X)$ to estimate an estimate of the unknown parameters.

 To implement the SIMEX estimation procedure,  we preselect a finite sequence of $\lambda$-values $\lambda_1, \lambda_2, \ldots, \lambda_M$ from an interval $\Lambda=[\lambda_1, \lambda_M]$, and a sufficiently large positive integer $B$. Often times equally spaced $\lambda$-values with $\lambda_1=0$ and $\lambda_M=2$ are used. Then we follow the three steps below to estimate $\theta$.\vskip 0.2in

 \begin{minipage}{0.95\textwidth}

   {\bf Simulation:} For $\lambda=\lambda_1$, independently generate $B$ sets of normal random numbers of size $n$ from $N_p(0,\Sigma_u)$. In particular, for the $b$-th set, generate $V_{ib}$ i.i.d. $\sim N_p(0,\Sigma_u)$, and calculate
          $
            W_{ib}(\lambda)=X_i+U_i+\sqrt{\lambda_1}V_{ib}=W_i+\sqrt{\lambda_1}V_{ib},\, i=1,\ldots,n.
          $
 \end{minipage} \vskip 0.1in

 \begin{minipage}{0.95\textwidth}

  {\bf Estimation:} For each $b=1,2,\ldots, B$, calculate
          $
            \hat\theta_b(\lambda_1)=\mbox{argmax}_\theta Q_n(\theta,\lambda_1)
          $
         where
           \begin{equation}\label{eq2.2}
             Q_n(\theta,\lambda)=\frac{1}{nh}\sum_{i=1}^nK\left(\frac{Y_i-m(W_{ib}(\lambda),\theta)}{h}\right).
           \end{equation}
         and the average
           $
             \hat\theta(\lambda_1)=B^{-1}\sum_{b=1}^B\hat\theta_b(\lambda_1).
           $
         Iterate the Simulation-Estimation steps for $\lambda=\lambda_2, \ldots, \lambda_M$, and obtain the sequence $\hat\theta(\lambda_1)$, $\ldots,$ $\hat\theta(\lambda_M)$.\vskip 0.1in
   \end{minipage}\vskip 0.1in

  \begin{minipage}{0.95\textwidth}
   {\bf Extrapolation:} Identify a trend of $\hat\theta(\lambda)$ versus $\lambda$, then extrapolate the trend to $\lambda=-1$ to obtain the SIMEX estimate $\hat\theta(-1)$.
 \end{minipage}\vskip 0.2in

  Motivations and theoretical justification of SIMEX algorithm can be found in the seminal papers by \cite{cook1994}, \cite{stefan1995} and \cite{carroll1996}. In general, the simulation and the estimation steps cause no trouble, however, extra caution should be paid in the extrapolation step, since in most cases, the exact extrapolation function is not available. Although three alternatives, such as the linear function $a+b\lambda$, the quadratic function $a+b\lambda+c\lambda^2$ and the nonlinear function $a+c/(d+\lambda)$, are often recommended in literature, they are simply empirical suggestions, except for some special cases. To avoid this technical difficulty, instead of directly dealing with the issue, most research done in literature simply assumes the true extrapolation function to be known. See \cite{carroll1996} and \cite{yang2019} for more details.

 \section{Asymptotic Results of The SIMEX Estimator}\label{sec3}

 In this section, we shall justify the SIMEX algorithm proposed in Section \ref{sec2} works well in model (\ref{eq2.1}) by stating some large sample properties, including the consistency and asymptotic normality, of the proposed estimator of $\theta$. To begin with, for a kernel function $K$ and a sequence of vanishing positive numbers $h$, depending on the sample size, denote $K_h(t)=h^{-1}K(t/h)$, and define
    \begin{eqarray}
      \theta(\lambda, h)&=&\mbox{argmax}_\theta EK_h(Y-m(W(\lambda),\theta)), \label{eq3.1}\\
      \theta(\lambda)&=&\mbox{argmax}_\theta \lim_{h\to 0}EK_h(Y-m(W(\lambda),\theta)). \label{eq3.2}
    \end{eqarray}
 For some technical reasons, see the proofs presented in Appendix, we shall deliberately choose $K$ to be the standard normal density function. We denote the conditional density function of $\vep$ given $X=x$ as $g(\vep|X=x)$. The following is a list of technical conditions needed for the statement of the main results, as well as their proofs.

 \begin{itemize}
   \item[] {\bf C1.} $\dddot m(x,\theta)$ with respect to $\theta$ is continuous; $m'(x,\theta)$ is continuous with respect to $x$.
   \item[] {\bf C2.} For each $\lambda$, $E[g(t+m(W(\lambda),\theta)-m(X,\theta_0))|W(\lambda)]$ has up to third order continuous and bounded derivatives, and
         $$E\|\dot m(W(\lambda),\theta(\lambda))g'(m(W(\lambda),\theta(\lambda))-m(X,\theta_0))\|^2<\infty.$$
   \item[] {\bf C3.} $g'(0|X=x)=0$, $g''(0|X=x)<0$, $g^{(k)}(\vep|X=x)$ are continuous and bounded for $k=0,1,2,3$ for any $x$, and for all $\lambda\geq 0$,
                         $$
                           \frac{\partial^2}{\partial\theta\partial\theta^T}E\left[g(m(W(\lambda),\theta(\lambda))-m(X,\theta_0))|X\right]
                         $$
                     is negative definite.
   \item[] {\bf C4.} For each $\lambda$, for $n$ sufficiently large, the maximizer $\theta(\lambda,h)$ is unique, and is the solution of
         $$
           E\left[\frac{\partial K_h(Y-m(W(\lambda),\theta))}{\partial\theta}\right]=\frac{\partial EK_h(Y-m(W(\lambda),\theta))}{\partial\theta}=0.$$
   \item[] {\bf C5.} The bandwidth $h\to 0$, $nh^5\to\infty$ as $n\to\infty$.
 \end{itemize}

 The above conditions are mild and similar to those assumptions imposed for the linear model regression in \cite{Yao2014A}, but they are modified accordingly for the parametric and measurement error setup.

 We start with a theorem regarding the relationship between $\theta(\lambda,h)$ and $\theta(\lambda)$ defined in (\ref{eq3.1}) and (\ref{eq3.2}), respectively.

 \begin{thm}\label{thm1} For any fixed $\lambda\in [\lambda_1,\lambda_M]$,
    $
      \theta(\lambda,h)=\theta(\lambda)+D^{-1}(\lambda)C(\lambda)h^2+o(h^2),
    $
  where
    \begin{eqarray*}
      C(\lambda)&=&\frac{1}{2}E\int g'''(m(X+\tau v,\theta(\lambda))-m(X,\theta_0)|X)\dot m(X+\tau v,\theta(\lambda))\phi(v)dv,\\
      D(\lambda)&=&E\int g''(m(X+\tau v,\theta(\lambda))-m(X,\theta_0)|X)(\dot m(X+\tau v,\theta(\lambda)))^{\otimes 2}\phi(v)dv
    \end{eqarray*}
  and $\tau=\sqrt{1+\lambda}\Sigma_u^{1/2}$, and $\phi(v)$ is the density function of $p$-dimensional standard normal distribution.
 \end{thm}

The following theorem shows that the distance between $\hat\theta(\lambda)$ and $\theta(\lambda,h)$ vanishes as $n\to\infty$.

\begin{thm}\label{thm2} Suppose (C1)-(C5) holds. Then there exits a maximizer $\hat\theta(\lambda)$ such that
    $
      \|\hat\theta(\lambda)-\theta(\lambda,h)\|=O_p(a_n),
    $
where $a_n=h^2+(nh^3)^{-1/2}$.
\end{thm}

Denote
 $$
   \Lambda=(\lambda_1,\ldots,\lambda_M)^T,\quad \theta(\Lambda)=(\theta^T(\lambda_1),\ldots,\theta^T(\lambda_M))^T,\quad C(\Lambda)=(C^T(\lambda_1),\ldots,C^T(\lambda_M))^T
 $$
 $$
   D(\Lambda)=\mbox{diag}(D(\lambda_1),\ldots, D(\lambda_M)),\quad
   J(\Lambda)=\mbox{diag}\left(J(\lambda_{1}),\ldots, J(\lambda_{M})\right),
 $$
 and
 $$
   \Pi(\Lambda)=\mbox{diag}\left(
   \frac{1}{4B\sqrt{\pi}}E\bigg[\bigg(\dot m(W(\lambda_j),\theta(\lambda_j,h))\bigg)^{\otimes 2}f_{\lambda}(0|W(\lambda_j))\bigg]\right).
 $$
 The following theorem claims that $\hat\theta(\Lambda)$ is asymptotically multivariate normal.

\begin{thm}\label{thm3} Under the same conditions as in Theorem \ref{thm2},
    $$
     \sqrt{nh^{3}}(\hat\theta(\Lambda)-\theta(\Lambda)-D^{-1}(\Lambda)C(\Lambda)h^2+o(h^2))\Longrightarrow N(0, J^{-1}(\Lambda)\Pi(\Lambda)J^{-1}(\Lambda)).
    $$
\end{thm}

 To our surprise, the above theorem actually indicates that $\hat\theta(\lambda_1),\ldots, \hat\theta(\lambda_M)$ are asymptotically independent! which seems inconsistent with the results obtained in \cite{yang2019} in the single index regression setup. We have double checked some special cases, such as when the modal regression function is linear, and the measurement error is normal, and found out it is indeed the case.

 To further derive the large sample properties of the SIMEX estimator $\hat\theta_{\rm\tiny SIMEX}$ based on the above results, we have to know the form of the extrapolation function $\theta(\lambda)$. As we mentioned in Section 2, no explicit extrapolation function form is available except for some rare cases. To see this point, we note that $\theta(\lambda)$ is the solution of the following equation
    $$
      \frac{\partial}{\partial\theta} Eg(m(W(\lambda,\theta))-m(X,\theta_0))=0
    $$
 or
   $$
     E\int g'(m(X+u,\theta)-m(X,\theta_0))\dot m(X+u,\theta)\exp\left(-\frac{u'\Sigma_u^{-1}u}{2(1+\lambda)}\right)du=0
   $$
 The justification of this statement can be found in the proof of Theorem $\ref{thm2}$ in Appendix.

 For illustration purpose, assume $\vep$ and $X$ are independent and standard normal, the modal regression function is linear, $m(x,\theta)=\theta x$, then simple calculation shows
  $$
    \lim_{h\to 0}EK_h(Y-m(W(\lambda),\theta))=\frac{1}{\sqrt{2\pi (1+(\theta-\theta_0)^2\sigma_x^2+(1+\lambda)\sigma_u^2\theta^2)}}
  $$
and $\theta(\lambda)$ is the solution of
  $$
    \frac{\partial (1+(\theta-\theta_0)^2\sigma_x^2+(1+\lambda)\sigma_u^2\theta^2)}{\partial\theta}=0,
  $$
which gives the exact extrapolation function
  $$
    \theta(\lambda)=\frac{\theta_0\sigma_x^2}{\sigma_x^2+(1+\lambda)\sigma_u^2}.
  $$
Clearly, the exact extrapolation function has the nonlinear form $a+c/(d+\lambda)$, and indeed $\theta(-1)=\theta_0$.

However, in real applications, the density functions of $\vep$ and $X$ are unknown, $m$ may have a complicated form, so there is no way to obtain a manageable form of $\theta(\lambda)$. So, in the following, we will adopt the strategy used in literature, simply assuming the extrapolation function has a parametric form. In the real application, estimated extrapolation function by fitting the pairs $(\lambda_j,\hat\theta(\lambda_j))$ should be used to approximate the true SIMEX estimator.

Suppose the true extrapolation function $\theta(\lambda)$ has the form $G(\lambda,\Gamma_0)$, which is twice continuously differentiable with respect to the unknown parameter $\Gamma_0\in\R^d$ for some positive integer $d$. We will estimate $\Gamma_0$ by minimizing the least squares criterion $\|\hat\theta(\Lambda)-G(\Lambda,\Gamma)\|^2$, where
  $$
    G(\Lambda,\Gamma)=[G^T(\lambda_1,\Gamma), G^T(\lambda_2,\Gamma),\ldots,G^T(\lambda_M,\Gamma)]^T_{qm\times 1}
  $$
 or solving the equation $\dot G^T(\Lambda,\Gamma)(\hat\theta(\Lambda)-G(\Lambda,\Gamma))=0$, where
  $$
    \dot G(\Lambda,\Gamma)=[\dot G^T(\lambda_1,\Gamma), \dot G^T(\lambda_2,\Gamma),\ldots,\dot G^T(\lambda_M,\Gamma)]^T_{qm\times d},
  $$
 and
  $$
    \dot G(\lambda_j,\Gamma)=\left(\frac{\partial G_k(\lambda_j,\Gamma)}{\partial\gamma_l}\right)_{q\times d}.
  $$
 Suppose $\hat\Gamma$ is the solution, then by Taylor expansion, we have
 \begin{eqarray*}
    &&0=\dot G^T(\Lambda,\hat\Gamma)(\hat\theta(\Lambda)-G(\Lambda,\hat\Gamma))\\
   &=&\dot G^T(\Lambda,\Gamma_0)(\hat\theta(\Lambda)-G(\Lambda,\Gamma_0))
  +\left[T(\Lambda,\tilde\Gamma)
  -\dot G^T(\Lambda,\tilde\Gamma)\dot G(\Lambda,\tilde\Gamma)
  \right](\hat\Gamma-\Gamma_0),
 \end{eqarray*}
 where $\tilde\Gamma$ is between $\hat\Gamma$ and $\Gamma_0$ and
   $$
     T(\Lambda,\Gamma)=\sum_{j=1}^M\sum_{k=1}^q \begin{pmatrix}\frac{\partial G_k(\lambda_j,\Gamma)}{\partial\gamma_1\partial\Gamma^T}(\hat\theta_k(\lambda_j)-G_k(\lambda_j,\Gamma))\\
     \vdots \\
     \frac{\partial G_k(\lambda_j,\Gamma)}{\partial\gamma_d\partial\Gamma^T}(\hat\theta_k(\lambda_j)-G_k(\lambda_j,\Gamma))
     \end{pmatrix}_{d\times d}
   $$

 The consistency of $\hat\Gamma$ to $\Gamma_0$ implies that
  \begin{eqarray*}
   \sqrt{nh^3}\left[\dot G^T(\Lambda,\Gamma_0)\dot G(\Lambda,\Gamma_0)+o_p(1)\right](\hat\Gamma-\Gamma_0)=\sqrt{nh^3}\dot G^T(\Lambda,\Gamma_0)(\hat\theta(\Lambda)-\theta(\Lambda)).
  \end{eqarray*}
Therefore, denote $H(\Lambda)=\dot G^T(\Lambda,\Gamma_0)\dot G(\Lambda,\Gamma_0)$,
  \begin{eqarray*}
   &&\sqrt{nh^3}H(\Lambda)\left[\hat\Gamma-\Gamma_0-H^{-1}(\Lambda)\dot G^T(\Lambda,\Gamma_0)(
   D^{-1}(\Lambda)C(\Lambda)h^2+o(h^2))\right]\\
   &=&\sqrt{nh^3}\dot G^T(\Lambda,\Gamma_0)(\hat\theta(\Lambda)-\theta(\Lambda))-\sqrt{nh^3}\dot G^T(\Lambda,\Gamma_0)(D^{-1}(\Lambda)C(\Lambda)h^2+o(h^2))\\
   &=&\sqrt{nh^3}\dot G^T(\Lambda,\Gamma_0)(\hat\theta(\Lambda)-\theta(\Lambda)-D^{-1}(\Lambda)C(\Lambda)h^2+o(h^2)).
  \end{eqarray*}
This implies that, from Theorem \ref{thm3}, if $H(\Lambda)$ is nonsingular,
  \begin{equation}\label{eq3.3}
    \sqrt{nh^3}\left[\hat\Gamma-\Gamma_0-S(\Lambda)h^2+o(h^2)\right]
    \Longrightarrow N(0,\Sigma(\Lambda)).
  \end{equation}
with
  \begin{eqarray*}\Sigma(\Lambda)&=&H^{-1}(\Lambda)\dot G^T(\Lambda,\Gamma_0)J^{-1}(\Lambda)\Pi(\Lambda)J^{-1}(\Lambda)\dot G(\Lambda,\Gamma_0)H^{-1}(\Lambda),\\
  S(\Lambda)&=&H^{-1}(\Lambda)\dot G^T(\Lambda,\Gamma_0)D^{-1}(\Lambda)C(\Lambda).
  \end{eqarray*}
Note that the SIMEX estimate $\hat\theta_{\rm\tiny SIMEX}$ is defined as $\hat\theta_{\rm\tiny SIMEX}=G(-1,\hat\Gamma)$, also note that $G(-1,\Gamma_0)=\theta_0$, so by Taylor expansion again, $ \hat\theta_{\rm\tiny SIMEX}-\theta_0=\dot G(-1, \tilde\Gamma)(\hat\Gamma-\Gamma_0)$, together with the asymptotic result (\ref{eq3.3}), we have the following theorem.

\begin{thm}\label{thm4} In addition to the conditions in Theorem \ref{thm3}, if we further assume that the true extrapolation function is $G(\lambda,\Gamma)$,  $H(\Lambda)$ is nonsingular and $nh^7=O(1)$, then
  \begin{eqarray*}
  \sqrt{nh^{3}}(\hat{\theta}_{\tiny SIMEX}-\theta_{0}-\dot G(-1,\Gamma_0) S(\Gamma_0)h^{2}+o(h^2))\Longrightarrow N(0,\dot G(-1,\Gamma_0)\Sigma(\Lambda)\dot G^T(-1,\Gamma_0)).
  \end{eqarray*}
\end{thm}

From Theorem \ref{thm4}, we can see that the asymptotic mean squared error of $\hat\theta_{\tiny SIMEX}$ is $h^4\|\dot G(-1,\Gamma_0) S(\Gamma_0)\|^{2}+(nh^3)^{-1}\mbox{Trace}( G(-1,\Gamma_0)\Sigma(\Lambda)\dot G^T(-1,\Gamma_0))$, thus an asymptotic optimal bandwidth can be obtained by minimizing the asymptotic mean squared error,
   $$
     h_{\tiny opt}=\left[\frac{3\mbox{Trace}( G(-1,\Gamma_0)\Sigma(\Lambda)\dot G^T(-1,\Gamma_0))}{4n\|\dot G(-1,\Gamma_0) S(\Gamma_0)\|^{2}}\right]^{1/7}.
   $$
However, $h_{\tiny opt}$ depends on some unknown quantities, thus it cannot be applied directly. Certain approximations are needed. See \cite{Yao2014A} for a relevant discussion in the linear modal case.

\section{Numerical Study}

 To evaluate the finite performance of the proposed SIMEX estimator of the modal regression coefficient, in this section, we shall conduct a simulation study. Note that the estimation step in the SIMEX algorithm described in Section 2 requires the maximization of $Q_n(\theta,\lambda)$ with respect to $\theta$ for each $\lambda$, similar to linear modal regression case, there is no explicit solution. Instead, we can define a similar EM algorithm as in \cite{Yao2014A}. For the sake of completeness, the main steps are listed in the following. \vskip 0.1in

  \begin{minipage}{0.95\textwidth}
    {\bf E-Step:} For an initial value $\theta^{(0)}$, calculate the weights $\pi(j|\theta^{(0)}), j=1,2,\ldots,n$
        $$
          \pi(j|\theta^{(0)})=\frac{\phi_h\left(Y_j-m(W_{jb}(\lambda),\theta^{(0)})\right)}{\sum_{i=1}^n\phi_h\left(Y_i-m(W_{ib}(\lambda),\theta^{(0)})\right)}.
        $$
   \end{minipage}\vskip 0.2in

  \begin{minipage}{0.95\textwidth}

    {\bf M-Step:} Maximize the new target function
        \begin{equation}\label{eq4.1}
          \sum_{j=1}^n \pi(j|\theta^{(0)})\log\phi_h(Y_j-m(W_{jb}(\lambda),\theta))
        \end{equation}
     with respect to $\theta$.\vskip 0.05in

    {\bf Iteration Step:} Using the maximizer obtained in the M-step as the new initial value, and iterate the above E- and M-step until some convergence criterion is met.
  \end{minipage}\vskip 0.2in

 It is easy to see, to maximize (\ref{eq4.1}) is equivalent to minimize
    $$
                \sum_{j=1}^n \pi(j|\theta^{(0)})(Y_j-m(W_{jb}(\lambda),\theta))^2.
    $$
 However, for nonlinear function $m$, the minimizer does not have a close form and numerical solution should be sought.

 This EM algorithm is useful when the dimension $q$ of $\theta$ is high. If $q$ is relatively small, some functions from existing R package can be used to derive the solution.

 The data used in the simulation study are generated from the following modal regression model
   $
     Y=\alpha\exp(\beta X)+\sigma(X)\varepsilon,
   $
where $X\sim U(0,1)$, $\varepsilon\sim 0.5N(-1,2.5^2)+0.5N(1,0.5^2)$, $X$ and $\varepsilon$ are independent, and $\sigma(X)=\gamma\exp(\beta X)$. Since $E\varepsilon=0$, $Mod(\varepsilon)=1$, and $Med(\varepsilon)=0.67$, so it is easy to see that $E[Y|X]=\alpha\exp(\beta X)$,
Mode$[Y|X]=(\alpha+\gamma)\exp(\beta X)$, Median$[Y|X]=(\alpha+0.67\gamma)\exp(\beta X)$. This model is similar to the one used in \cite{Yao2014A} except for the regression function. In the simulation study, we choose the true values of the parameters to be $\alpha=\beta=\gamma=1$. Therefore, the true conditional mean, mode and median are $E[Y|X]=\exp(X)$, Mode$[Y|X]=2\exp(X)$ and Median$[Y|X]=1.67\exp(X)$, respectively. For the measurement error model $W=X+U$, we choose $U\sim N(0, \sigma_u^2)$ with $\sigma_u^2=0.01, 0.02, 0.04$. Note that the variance of $X$ is $1/12\approx 0.08$, so the noise-to-signal ratio is roughly $12.5\%, 25\%$ and $50\%$ respectively. Two sample sizes of $n=200$ and $400$ are used in the simulation study. In each scenario, the simulation is repeated 100 times, the mean, bias and mean squared errors (MSE) are computed to evaluated the finite sample performance of the estimation procedures. For all SIMEX related algorithm, $B=50$, and the $\lambda$-values are 10 equally spaced points from $[0,2]$. To evaluate the effect of the bandwidth on the estimate, we choose $h=cn^{-1/7}$ and $c=0.5, 0.8, 1, 1.2$.

In addition to the proposed SIMEX modal regression procedure (S-Modal),  we also consider the other five methods for estimating the mean or modal regression parameters:
 \begin{itemize}
   \item Naive Mean Regression Based on LSE (N-Mean). The target function to minimize is
      $\sum_{j=1}^n [Y_i-\alpha\exp(\beta W_i)]^2$.
   \item SIMEX mean regression based on LSE (S-Mean). The steps for implementing the SIMEX mean regression are exactly the same as in the classic SIMEX mean regression based LSE. In particular, in the estimation step, the following target function is minimized
        $\sum_{j=1}^n [Y_i-\alpha\exp(\beta W_{ik}(\lambda))]^2$.
   \item SIMEX M-estimate based on Huber's weight $\rho$-function (S-Huber). Huber's weight family of functions is defined as $\rho(x)=0.5x^2$ if $|x|\leq c$, and $c|x|-0.5c^2$ if $|x|>c$.  The constant $c$ for $95\%$ efficiency of the regression estimator is $1.345\sigma$, where $\sigma$ is the standard deviation of the errors.  Therefore, we obtain the estimate by minimizing the following target function
        $\sum_{i=1}^n\rho(Y_i-m(Z_i(\lambda),\theta))$.
   \item SIMEX median regression estimate (S-Median). The steps for implementing the SIMEX median regression procedure are the same as in the SIMEX mean regression based LSE, the only difference is to replace the target function to
        $\sum_{j=1}^n |Y_i-\alpha\exp(\beta W_{ik}(\lambda))|$.
  \item Naive-Modal regression estimate (N-Modal). Directly minimizing (\ref{eq2.1.1}) with $X_i$ replaced by $W_i$.
  \end{itemize}

These six methods can be classified into three groups. The first group consists of the Naive mean regression, the SIMEX mean regression and the SIMEX M-estimate, they are used to fit the mean regression function $\alpha\exp(\beta X)$; the second group includes the SIMEX median regression, which is used for estimating the median regression function $(\alpha+0.67\gamma)\exp(\beta X)$; the third group, consisting of the proposed SIMEX modal regression and the naive modal regression, is used for fit the modal regression function $(\alpha+\gamma)\exp(\beta X)$. The comparison should be made within each group, but we can assess the robustness cross different procedures. To obtain estimates of the unknown parameters, for S-Median and S-Huber, we use the function \texttt{optim} from R package \texttt{MASS}, and for other methods, function \texttt{nlrobe} from R package \texttt{robustbase} is used.

Simulation results are summarized in the Table \ref{table1}-\ref{table6} for $c=0.8$.  For the two mean and modal regression methods, it is not surprising to see the S-Mean method and the proposed modal regression procedure performs better in reducing the bias than the naive mean and modal regressions, which simply ignore the measurement error by treating the error-prone variable as the true predictor, however, the variances from the SIMEX procedures are relatively larger than their naive counterparts. The SIMEX M-estimate based Huber's weight function show noticeable biases in all cases, while the S-Median fits the median regression function very well. Also, one can notice that the estimates are getting worse when the variance of the measurement error is getting larger, and performance improves when the sample size gets larger.

 \begin{table}
  \centering
   \begin{tabular}{|c|c|rrrrrr|}
    \hline\hline
         &       &  N-Mean      & S-Mean     &  S-Huber & S-Median   & S-Modal    &  N-Modal \\ \hline
         & Mean  &  1.084       &    1.020   &   1.152  &  1.595     &   1.996    &   2.133  \\
$\alpha$ & Bias  &  0.084       &    0.020   &   0.152  & -0.075     &  -0.004    &   0.133  \\
%a:variance &    &  0.115       &    0.132   &   0.088  &  0.080     &   0.128    &   0.072  \\
         & MSE   &  0.122       &    0.133   &   0.111  &  0.086     &   0.128    &   0.090  \\ \hline
         & Mean  &  0.929       &    1.043   &   1.247  &  0.939     &   0.940    &   0.706  \\
$\beta$  & Bias  & -0.071       &    0.043   &   0.247  & -0.060     &  -0.060    &  -0.294  \\
%b:variance &    &  0.306       &    0.439   &   0.136  &  0.072     &   0.149    &   0.047  \\
         & MSE   &  0.311       &    0.441   &   0.197  &  0.076     &   0.152    &   0.133  \\ \hline\hline
   \end{tabular}
   \caption{$n=200, \sigma_u^2=0.01$}\label{table1}
 \end{table}

 \begin{table}
  \centering
   \begin{tabular}{|c|c|rrrrrr|}
    \hline\hline
         &       &  N-Mean      & S-Mean     &  S-Huber& S-Median    & S-Modal &  N-Modal \\ \hline
         & Mean  &  1.134 &  1.030 &  1.170 &   1.606 &   2.010 &   2.232  \\
$\alpha$ & Bias  &  0.134 &  0.030 &  0.170 &  -0.064 &   0.010 &   0.232  \\
%a:variance &    &  0.117 &  0.146 &  0.095 &   0.103 &   0.168 &   0.100  \\
         & MSE   &  0.135 &  0.147 &  0.124 &   0.107 &   0.168 &   0.153  \\ \hline
         & Mean  &  0.838 &  1.016 &  1.208 &   0.902 &   0.910 &   0.603  \\
$\beta$  & Bias  & -0.162 &  0.016 &  0.208 &  -0.098 &  -0.090 &  -0.397  \\
%b:variance &    &  0.247 &  0.409 &  0.135 &   0.053 &   0.188 &   0.051  \\
         & MSE   &  0.274 &  0.409 &  0.178 &   0.063 &   0.196 &   0.208  \\ \hline\hline
   \end{tabular}
   \caption{$n=200, \sigma_u^2=0.02$}
   \label{table2}
 \end{table}

 \begin{table}
  \centering
   \begin{tabular}{|c|c|rrrrrr|}
    \hline\hline
         &       &  N-Mean & S-Mean   &  S-Huber & S-Median  & S-Modal &  N-Modal \\ \hline
         & Mean  &   1.202 &   1.053  &  1.199   &  1.505    &  2.071  &  2.314  \\
$\alpha$ & Bias  &   0.202 &   0.053  &  0.199   & -0.165    &  0.071  &  0.314  \\
%a:variance &    &   0.111 &   0.154  &  0.097   &  0.214    &  0.204  &  0.113  \\
         & MSE   &   0.152 &   0.157  &  0.137   &  0.241    &  0.209  &  0.212  \\ \hline
         & Mean  &   0.710 &   0.948  &  1.123   &  0.924    &  0.837  &  0.519  \\
$\beta$  & Bias  &  -0.290 &  -0.052  &  0.123   & -0.076    & -0.163  & -0.481  \\
%b:variance &    &   0.194 &   0.384  &  0.135   &  0.072    &  0.204  &  0.048  \\
         & MSE   &   0.278 &   0.387  &  0.150   &  0.078    &  0.231  &  0.279  \\ \hline\hline
   \end{tabular}
   \caption{$n=200, \sigma_u^2=0.04$}
   \label{table3}
 \end{table}

 \begin{table}
  \centering
   \begin{tabular}{|c|c|rrrrrr|}
    \hline\hline
         &       &  N-Mean    & S-Mean    &  S-Huber & S-Median  & S-Modal &  N-Modal \\ \hline
         & Mean  &   1.102    &   1.040   &  1.139   &   1.568   & 2.034   &   2.155    \\
$\alpha$ & Bias  &   0.102    &   0.040   &  0.139   &  -0.102   & 0.034   &   0.155    \\
%a:variance &    &   0.069    &   0.087   &  0.051   &   0.024   & 0.082   &   0.046    \\
         & MSE   &   0.079    &   0.089   &  0.070   &   0.035   & 0.083   &   0.070    \\ \hline
         & Mean  &   0.841    &   0.952   &  1.246   &   0.900   & 0.937   &   0.737    \\
$\beta$  & Bias  &  -0.159    &  -0.048   &  0.246   &  -0.100-  & 0.063   &  -0.263    \\
%b:variance &    &   0.182    &   0.258   &  0.085   &   0.035   & 0.118   &   0.036    \\
         & MSE   &   0.207    &   0.260   &  0.146   &   0.045   & 0.122   &   0.105    \\ \hline\hline
   \end{tabular}
   \caption{$n=400, \sigma_u^2=0.01$}
   \label{table4}
 \end{table}
 
 \begin{table}
  \centering
   \begin{tabular}{|c|c|rrrrrr|}
    \hline\hline
         &       &  N-Mean  & S-Mean    &  S-Huber & S-Median   & S-Modal  &  N-Modal \\ \hline
         & Mean  &   1.151  &    1.052  &   1.160  &     1.619  &   2.017  &    2.253  \\
$\alpha$ & Bias  &   0.151  &    0.052  &   0.160  &    -0.051  &   0.017  &    0.253  \\
%a:variance &    &   0.065  &    0.095  &   0.052  &     0.075  &   0.101  &    0.055  \\
         & MSE   &   0.088  &    0.098  &   0.078  &     0.077  &   0.101  &    0.119  \\ \hline
         & Mean  &   0.755  &    0.924  &   1.204  &     0.869  &   0.961  &    0.615  \\
$\beta$  & Bias  &  -0.245  &   -0.076  &   0.204  &    -0.131  &  -0.039  &   -0.385  \\
%b:variance &    &   0.156  &    0.266  &   0.085  &     0.034  &   0.101  &    0.039  \\
         & MSE   &   0.216  &    0.271  &   0.127  &     0.051  &   0.102  &    0.187  \\ \hline\hline
   \end{tabular}
   \caption{$n=400, \sigma_u^2=0.02$}
 \end{table}

 \begin{table}
  \centering
   \begin{tabular}{|c|c|rrrrrr|}
    \hline\hline
         &       &  N-Mean    & S-Mean      &  S-Huber   &S-Median  &  S-Modal &  N-Modal \\ \hline
         & Mean  &    1.227    &   1.090   &  1.214  &  1.437    &   1.996  &    2.358 \\
$\alpha$ & Bias  &    0.227    &   0.090   &  0.214  & -0.233    &  -0.004  &    0.358 \\
%a:variance &    &    0.060    &   0.102   &  0.055  &  0.252    &   0.137  &    0.069 \\
         & MSE   &    0.112    &   0.110   &  0.101  &  0.306    &   0.137  &    0.197 \\ \hline
         & Mean  &    0.629    &   0.846   &  1.099  &  0.919    &   0.947  &    0.497 \\
$\beta$  & Bias  &   -0.371    &  -0.154   &  0.099  & -0.081    &  -0.053  &   -0.503 \\
%b:variance &    &    0.121    &   0.250   &  0.081  &  0.071    &   0.137  &    0.034 \\
         & MSE   &    0.258    &   0.273   &  0.091  &  0.078    &   0.140  &    0.286 \\ \hline\hline
   \end{tabular}
   \caption{$n=400, \sigma_u^2=0.04$}
   \label{table6}
 \end{table}

The simulation results for other $c$-values are also conducted. Similar patterns are obtained, which indicate the proposed estimation procedure is stable, and for the sake of brevity, the corresponding simulation results are omitted.

\section{Appendix}

 \def\pdm{\frac{\partial m(W(\lambda),\theta)}{\partial\theta}}
 \def\pdmi{\frac{\partial m(W_i(\lambda),\theta)}{\partial\theta}}

 This appendix contains the proofs of all the main results from Section \ref{sec3}. For the sake of simplicity, we only present the proof of univariate $X$, the extension to $p$-dimensional covariates is straightforward, except for some notational complexity. Thus $\tau=\sqrt{1+\lambda}\sigma_u$.  \vskip 0.1in

 \begin{proof}[The proof of Theorem \ref{thm1}]

  First, let us show that $\theta(\lambda,h)\to\theta(\lambda)$ as $h\to 0$.
  Denote
    \begin{eqarray*}
     g_m(x,v,u;\theta,\theta_0)&=&g(m(x+\tau v,\theta)-m(x,\theta_{0})+u|x),\\
     g_m'(x,v,u;\theta,\theta_0)&=&\partial g(t|x)/\partial t\Big|_{t=m(x+\tau v,\theta)-m(x,\theta_{0})+u}.
    \end{eqarray*}
  $g_m''$ and $g_m'''$ are similarly defined.
  Integrating by parts, we have
    \begin{eqarray*}
     &&EQ_h(\theta,\lambda)= % E\int \phi_h(Y-m(X+\tau v,\theta))\phi(v)dv\\
                          \frac{1}{h}E\iint \phi\bigg(\frac{\vep+m(X,\theta_{0})-m(X+\tau v,\theta)}{h}\bigg)g(\vep|X)\phi(v)dvd\vep\\
     &=&E\iint \phi(u)g_m(X,v,hu;\theta,\theta_0)\phi(v)dvdu=\iiint \phi(u)g_m(x,v,hu;\theta,\theta_0)\phi(v)f(x)dvdudx.
    \end{eqarray*}
  Therefore,
    \begin{eqarray*}
      &&  |EQ_h(\theta,\lambda)-\lim_{h\to 0}EQ_h(\theta,\lambda)|\\
      &\leq &\iiint \phi(u)|g_m(x,v,hu;\theta,\theta_0)-g_m(x,v,0;\theta,\theta_0)|\phi(v)f(x)dvdudx\\
      &=& h \iiint \phi(u)\left|g_m'(x,v,\tilde v;\theta,\theta_0)\right||u|\phi(v)f(x)dvdudx.
    \end{eqarray*}
  By the boundedness of the partial derivative of $g$, we can easily see that
    $$
      \sup_{\theta\in\Theta} |EQ_h(\theta,\lambda)-\lim_{h\to 0}EQ_h(\theta,\lambda)|=o(1).
    $$
  This, together with the uniqueness of the minimizer of $\lim_{h\to 0}EQ_h(\theta,\lambda)$, implies that $\theta(\lambda,h)\to\theta(\lambda)$ as $h\to 0$.

  Note that $\theta(\lambda, h)$ is the solution of (\ref{eq3.1}), it satisfies
    $$
      \partial EQ_h(\theta,\lambda)/\partial\theta|_{\theta(\lambda,h)}=0.
    $$

  By Taylor expansion, we have
  \begin{eqarray*}
    0&=&\iiint \phi(u)g_m'(x,v,hu;\theta(\lambda,h),\theta_0)\dot m(x+\tau v,\theta(\lambda,h))\phi(v)f(x)dvdudx\\
     &=&\iint g_m'(x,v,0;\theta(\lambda,h),\theta_0)\dot m(x+\tau v,\theta(\lambda,h))\phi(v)f(x)dvdx\\
     & &+\frac{h^2}{2}\iiint u^2\phi(u)
     g_m'''(x,v,\tilde u;\theta(\lambda,h),\theta_0)\dot m(x+\tau v,\theta(\lambda,h))
      \phi(v)f(x)dvdudx\\
     &=&\iint g_m'(x,v,0;\theta(\lambda,h),\theta_0)\dot m(x+\tau v,\theta(\lambda,h))\phi(v)f(x)dvdx\\
     && +\iint g_m''(x,v,0;\theta(\lambda,h),\theta_0)
     \dot m(x+\tau v,\tilde\theta(\lambda))
     \dot m^T(x+\tau v,\theta(\lambda,h))
        \phi(v)f(x)dvdx\cdot\\
     &&\hskip 2in    (\theta(\lambda,h)-\theta(\lambda))\\
     && +\frac{h^2}{2}\iiint u^2\phi(u)g_m'''(x,v,\tilde u;\theta(\lambda,h),\theta_0)\dot m(x+\tau v,\theta(\lambda,h))
      \phi(v)f(x)dvdudx.
  \end{eqarray*}
  By the definition of $\theta(\lambda)$, the first term on the right hand side of the last equality is $0$. This, together with the claim $\theta(\lambda,h)\to\theta(\lambda)$ we just shown, implies
   \begin{eqarray*}
     0&=&\iint g_m''(x,v,0;\theta(\lambda,h),\theta_0)
     (\dot m(x+\tau v,\theta(\lambda)))^{\otimes2}
        \phi(v)f(x)dvdx
        (\theta(\lambda,h)-\theta(\lambda))(1+o(1))\\
     && +\frac{h^2}{2}\iint g_m'''(x,v,0;\theta(\lambda,h),\theta_0)\dot m(x+\tau v,\theta(\lambda))
      \phi(v)f(x)dvdx(1+o(1))
  \end{eqarray*}
 which indeed is the conclusion of Theorem \ref{thm1}.
 \end{proof}

 \begin{proof}[The proof of Theorem \ref{thm2}]
   It suffices to show that for an arbitrary number $\eta\in[0,1)$, there exists an sufficiently large number $C$ such that
     \begin{equation}\label{eq5.1}
       P\left\{\sup_{\|\mu\|=C}Q_n(\theta(\lambda,h)+a_n\mu)<Q_n(\theta(\lambda,h))\right\}\geq 1-\eta
     \end{equation}
   for $a_n=(nh^3)^{-1/2}+h^2$.
   Using Taylor expansion, we have
   \begin{eqarray}
     & &  Q_n(\theta(\lambda,h)+a_n\mu)-Q_n(\theta(\lambda,h))\nonumber\\
     &=& a_n\mu^T\pd+\frac{a_n^2}{2}\mu^T\pdd\mu+\frac{a_n^3}{6}L_n(\theta^*(\lambda,h),\mu),\label{eq5.100}
   \end{eqarray}
   where $\theta^*(\lambda,h)$ is between $\theta(\lambda,h)$ and $\theta(\lambda,h)+a_n\mu$, and
     $$
       L_n(\theta(\lambda,h),\mu)=\mu^T\left(\mu^T\frac{\partial^3Q_n(\theta(\lambda,h))}{\partial\theta\partial\theta^T\partial\theta_1}\mu,
       \cdots, \mu^T\frac{\partial^3Q_n(\theta(\lambda,h))}{\partial\theta\partial\theta^T\partial\theta_p}\mu\right)^T.
     $$

   By the definition of $\theta(\lambda, h)$ and (C4), we have $E\partial Q_n(\theta(\lambda,h),\lambda)/\partial\theta=0$. Now, we calculate the variance of $\partial Q_n(\theta(\lambda,h),\lambda)/\partial\theta$. Note that for standard normal kernel $\phi$,
     \begin{equation}\label{eq5.2}
       \phi_h'(t)=-\frac{t}{h^3}\phi\left(\frac{t}{h}\right),\,
       \phi_h''(t)=\frac{1}{h^3}\left(\frac{t^2}{h^2}-1\right)\phi\left(\frac{t}{h}\right),\,
       \phi_h'''(t)=\frac{1}{h^4}\left(\frac{3t}{h}-\frac{t^3}{h^3}\right)\phi\left(\frac{t}{h}\right).
     \end{equation}
   Denote $\bW_b(\lambda)=(W_{1b}(\lambda),\ldots,W_{bn}(\lambda))'$ and
   $\varepsilon_{ib}(\lambda)=Y_{i}-m(W_{ib}(\lambda),\theta)$.
   We have
    \begin{eqarray*}
      \frac{\partial Q_{n}(\theta(\lambda,h))}{\partial\theta}
      &=&-\frac{1}{n}\sum\limits^{n}_{i=1}\phi'_{h}(\varepsilon_{ib}(\lambda))\dot m(W_{ib}(\lambda),\theta),\\
      \frac{\partial^{2} Q_{n}(\theta(\lambda,h))}{\partial\theta\partial\theta^{T}}
      &=& \frac{1}{n}\sum\limits^{n}_{i=1}\phi''_{h}(\varepsilon_{ib}(\lambda))(\dot m(W_{ib}(\lambda),\theta))^{\otimes 2}-\frac{1}{n}\sum\limits^{n}_{i=1}\phi'_{h}(\varepsilon_{ib}(\lambda))\ddot m(W_{ib}(\lambda),\theta).
    \end{eqarray*}
   and
     \begin{eqarray*}
        L_n(\theta,\mu)&=&-\frac{1}{n}\sum\limits^{n}_{i=1}\phi'''_{h}(\varepsilon_{ib}(\lambda))\bigg(\mu^{T}\dot m(W_{ib}(\lambda),\theta)\bigg)^{3}\\
        &&+\frac{3}{n}\sum\limits^{n}_{i=1}\phi''_{h}(\varepsilon_{ib}(\lambda))\mu^{T}\ddot m(W_{ib}(\lambda),\theta)\mu\mu^{T}\dot m(W_{ib}(\lambda),\theta)-\frac{1}{n}\sum\limits^{n}_{i=1}
        \phi_h'(\varepsilon_{ib}(\lambda))F_i(\theta,\mu),
     \end{eqarray*}
   where
     \begin{eqarray*}
       F_i(\theta,\mu)=\mu^T\left(
                  \mu^T\frac{\partial^3m(W_{ib}(\lambda),\theta)}{\partial\theta\partial\theta^T\partial\theta_1}\mu, \cdots,
                  \mu^T\frac{\partial^3m(W_{ib}(\lambda),\theta)}{\partial\theta\partial\theta^T\partial\theta_p}\mu\right)^T.
     \end{eqarray*}

  The stochastic properties of $Q_n(\theta(\lambda,h)+a_n\mu)-Q_n(\theta(\lambda,h))$ depends on the partial derivatives of $Q_n(\theta)$. In the following we shall derive the asymptotic expansions of conditional expectations and covariance matrices of these partial derivatives. First, for the conditional expectation of $\partial Q_n(\theta)/\partial\theta$, we have
  \begin{eqarray*}
    &&E\left(\frac{\partial Q_{h}(\theta(\lambda,h))}{\partial\theta}|\bf W(\lambda)\right)= E\bigg(-\frac{1}{n}\sum\limits^{n}_{i=1}\phi'_{h}(\varepsilon_i(\lambda))\dot m(W_{ib}(\lambda),\theta(\lambda,h))| W_{i}(\lambda)\bigg)\\
    &=& -\frac{1}{n}\sum\limits^{n}_{i=1}\int\phi'_{h}(\varepsilon_i(\lambda))\dot m(W_{ib}(\lambda),\theta(\lambda,h))f_{\lambda}(\varepsilon_i(\lambda)|W_{i}(\lambda))d\varepsilon_i(\lambda)\\
    &=& -\frac{1}{n}\sum\limits^{n}_{i=1}\int\frac{\varepsilon_i(\lambda)}{h^{3}}\phi\bigg(\frac{\varepsilon_i(\lambda)}{h}\bigg)\dot m(W_{ib}(\lambda),\theta(\lambda,h))f_{\lambda}(\epsilon(\lambda)|W_{i}(\lambda))d\varepsilon_i(\lambda)\\
    &=&  -\frac{1}{n}\sum\limits^{n}_{i=1}\int\frac{1}{h}t\phi(t)\dot m(W_{ib}(\lambda),\theta(\lambda,h))(f_{\lambda}(0|W_i(\lambda))+f'_{\lambda} (0|W_{i}(\lambda))ht+O_{p}(h^3))dt
  \end{eqarray*}
  \begin{eqarray*}
    &=& -\frac{1}{n}\sum\limits^{n}_{i=1}\dot m(W_{ib}(\lambda),\theta(\lambda,h))f'_{\lambda}(0|W_{i}(\lambda))\int t^{2}\phi(t)dt\{1+O_{p}(h^3)\}\\
    &=& -\frac{1}{n}\sum\limits^{n}_{i=1}\dot m(W_{ib}(\lambda),\theta(\lambda,h))f'_{\lambda}(0|W_{i}(\lambda))+O_{p}(h^3)\\
    &=& -\frac{1}{n}\sum\limits^{n}_{i=1}\dot m(W_{ib}(\lambda),\theta(\lambda))f'_{\lambda}(0|W_{i}(\lambda))+O_{p}(h^3).
 \end{eqarray*}
 The last equality is a consequence of the continuity of $\dot m(x,\theta)$ with respect to $\theta$ and $\theta(\lambda,h)\to\theta(\lambda)$ by Theorem \ref{thm1}.

 In the following, we would like to argue that
    \begin{equation}\label{eq5.3}
      \frac{1}{n}\sum\limits^{n}_{i=1}\dot m(W_{ib}(\lambda),\theta(\lambda))f'_{\lambda}(0|W_{i}(\lambda))=O_p\left(\frac{1}{\sqrt{n}}\right).
    \end{equation}
 First, we claim that, for any $t$,
   \begin{equation}\label{eq5.4}
     f_\lambda'(t|W(\lambda))=E[g'(t+m(W(\lambda),\theta)-m(X,\theta_0)|X)|W(\lambda)].
   \end{equation}
 In fact, for any $t$,
  \begin{eqarray*}
    &   & P(Y-m(W(\lambda),\theta)\leq t| W(\lambda))=E[P(Y-m(W(\lambda),\theta)\leq t| W(\lambda),X)|W(\lambda)]\\
    & = & E\left[\int_{-\infty}^{t+m(W(\lambda),\theta)-m(X,\theta_0)}g(v|X)dv\Big|W(\lambda)\right]
  \end{eqarray*}
 which implies
   $$
     f_\lambda(t|W(\lambda))=E \left[g(t+m(W(\lambda),\theta)-m(X,\theta_0)|X)\Big| W(\lambda)\right].
   $$
 Therefore, (\ref{eq5.4}) can be obtained by taking derivative on the above equality with respect to $t$.
 In particular, if $t=0$, we have
   $$
     f_\lambda'(0|W(\lambda))=E \left[g'(m(W(\lambda),\theta)-m(X,\theta_0)|X)\Bigg| W(\lambda)\right].
   $$
 So,
  \begin{eqarray*}
   E\dot m(W(\lambda),\theta(\lambda))f'_{\lambda}(0|W(\lambda))
   &=&E\left(\dot m(W(\lambda),\theta(\lambda))E \left[g'(m(W(\lambda),\theta(\lambda))-m(X,\theta_0)|X)\Big| W(\lambda)\right]\right)\\
   &=&E\left(\dot m(W(\lambda),\theta(\lambda))g'(m(W(\lambda),\theta(\lambda))-m(X,\theta_0)|X)\right)\\
   &=&\frac{\partial}{\partial\theta}Eg(m(W(\lambda),\theta(\lambda))-m(X,\theta_0)|X)=0
  \end{eqarray*}
 by the definition of $\theta(\lambda)$. Therefore, the claim (\ref{eq5.3}) follows from the condition (C2).
 This, together with the condition $nh^3\to\infty$, implies
   $$
     E\left(\frac{\partial Q_{h}(\theta(\lambda,h))}{\partial\theta}|\bf W(\lambda)\right)=O_p(h^2).
   $$

 For the conditional covariance matrix of $\partial Q_{n}(\theta)/\partial\theta$ given $\bW_b(\lambda)$, we have
  \begin{eqarray*}
 &&\mathrm{Cov}\left(\frac{\partial Q_{h}(\theta(\lambda,h))}{\partial\theta}\bigg|\bW_b(\lambda)\right)
 = \mathrm{Cov}\left(-\frac{1}{n}\sum\limits^{n}_{i=1}\phi'_{h}(\varepsilon_i(\lambda))\dot m(W_{ib}(\lambda),\theta(\lambda,h))\bigg|W_{i}(\lambda)\right)\\
 &=& \frac{1}{n^{2}}\sum\limits^{n}_{i=1}\mathrm{Cov}\left\{\phi'_{h}(\varepsilon_i(\lambda))\dot m(W_{ib}(\lambda),\theta(\lambda,h))\bigg|W_{i}(\lambda)\right\}\\
 &=& \frac{1}{n^{2}}\sum\limits^{n}_{i=1}\bigg\{E\bigg((\phi'_{h}(\varepsilon_i(\lambda)))^{2}(\dot m(W_{ib}(\lambda),\theta(\lambda,h)))^{\otimes 2}\bigg|W_{i}(\lambda)\bigg)\\
 &&-\left(E\left(\phi'_{h}(\varepsilon_i(\lambda))\dot m(W_{ib}(\lambda),\theta(\lambda,h))\bigg| W_{i}(\lambda)\right)\right)^{\otimes 2}\bigg\}\\
 &=&\frac{1}{n^{2}}\sum\limits^{n}_{i=1}\bigg\{\int\bigg(\frac{\vep_i(\lambda)}{h^{3}}\bigg)^{2}\phi^{2}\bigg(\frac{\vep_i(\lambda)}{h}
 \bigg)\bigg(\dot m(W_{ib}(\lambda),\theta(\lambda,h))\bigg)^{\otimes 2}f_{\lambda}(\vep_i(\lambda)|W_{i}(\lambda))d\vep_i(\lambda)\\
 &&-\bigg(\int\bigg(\frac{\vep_i(\lambda)}{h^{3}}\bigg)\dot m(W_{ib}(\lambda),\theta(\lambda,h))\phi\bigg(\frac{\vep_i(\lambda)}{h}\bigg)f_{\lambda}(\vep_i(\lambda)|W_{i}(\lambda))d\vep_i(\lambda)\bigg)^{\otimes 2}\bigg\}\\
 &=&\frac{1}{n^{2}}\sum\limits^{n}_{i=1}\bigg\{\int\frac{1}{h^{3}}t^{2}\phi^{2}(t)\bigg(\dot m(W_{ib}(\lambda),\theta(\lambda,h))\bigg)^{\otimes 2}(f_{\lambda}(0|W_{i}(\lambda))+f'_{\lambda}(0|W_{i}(\lambda))ht+o_{p}(h))dt\\
 & &-\bigg(\int\frac{1}{h}t\phi(t)
 \dot m(W_{ib}(\lambda),\theta(\lambda,h))(f_{\lambda}(0|W_{i}(\lambda))+f'_{\lambda}(0|W_{i}(\lambda))ht+o_{p}(h))dt\bigg)^{\otimes 2}\bigg\}\\
 &=&\frac{1}{n^{2}}\sum\limits^{n}_{i=1}\bigg\{\frac{1}{h^{3}}\bigg(\dot m(W_{ib}(\lambda),\theta(\lambda,h))\bigg)^{\otimes 2}f_{\lambda}(0|W_{i}(\lambda)){\int t^{2}\phi^{2}(t)dt}\\
 &&-\bigg(\dot m(W_{ib}(\lambda),\theta(\lambda,h))f'_{\lambda}(0|W_{i}(\lambda))\bigg)^{\otimes 2}\bigg\}\{1+o_{p}(1)\}\\
 &=&\bigg\{\frac{1}{4\sqrt{\pi}nh^{3}}\cdot\frac{1}{n}\sum\limits^{n}_{i=1}\bigg(\bigg(\dot m(W_{ib}(\lambda),\theta(\lambda,h))\bigg)^{\otimes 2}f_{\lambda}(0|W_{i}(\lambda))\\
 &&-\frac{1}{n}\cdot\frac{1}{n}\sum\limits^{n}_{i=1}\bigg(\dot m(W_{ib}(\lambda),\theta(\lambda,h))f'_{\lambda}(0|W_{i}(\lambda))\bigg)^{\otimes 2}\bigg\}\{1+o_{p}(1)\}\\
 &=& \frac{1}{4\sqrt{\pi}nh^{3}}\cdot\frac{1}{n}\sum\limits^{n}_{i=1}\bigg(\bigg(\dot m(W_{ib}(\lambda),\theta(\lambda,h))\bigg)^{\otimes 2}f_{\lambda}(0|W_{i}(\lambda))\{1+o_{p}(1)\}
 \end{eqarray*}
 \begin{eqarray*}
 &=&\frac{1}{4\sqrt{\pi}nh^{3}}\cdot\frac{1}{n}\sum\limits^{n}_{i=1}\bigg[\bigg(\dot m(W_{ib}(\lambda),\theta(\lambda,h))\bigg)^{\otimes 2}f_{\lambda}(0|W_{i}(\lambda))\bigg]+o_{p}\bigg(\frac{1}{nh^{3}}\bigg)\\
 &=&\frac{1}{4\sqrt{\pi}nh^{3}}\cdot\frac{1}{n}\sum\limits^{n}_{i=1}\bigg[\bigg(\dot m(W_{ib}(\lambda),\theta(\lambda))\bigg)^{\otimes 2}f_{\lambda}(0|W_{i}(\lambda))\bigg]+o_{p}\bigg(\frac{1}{nh^{3}}\bigg)\\
 \end{eqarray*}
 Again, the last equality is a consequence of the continuity of $\dot m(x,\theta)$ with respect to $\theta$ and $\theta(\lambda,h)\to\theta(\lambda)$ by Theorem \ref{thm1}.

 Now, let's consider the asymptotic order of the second derivative of $Q_n(\theta(\lambda, h))$ with respect to $\theta$.
 \begin{eqarray*}
 &&E\bigg(\frac{\partial^{2}Q_{h}(\theta(\lambda,h))}{\partial\theta\partial\theta^{T}}\bigg|\bW(\lambda)\bigg)\\
 &=& E\bigg[\frac{1}{n}\sum\limits^{n}_{i=1}\phi''_{h}(\varepsilon_i(\lambda))(\dot m(W_{ib}(\lambda),\theta(\lambda,h)))^{\otimes 2}-\frac{1}{n}\sum\limits^{n}_{i=1}\phi'_{h}(\varepsilon_i(\lambda))\ddot m(W_{ib}(\lambda),\theta(\lambda,h))\bigg|
 W_{i}(\lambda)\bigg]\\
 &=& E\bigg[\frac{1}{n}\sum\limits^{n}_{i=1}\frac{1}{h^{3}}\bigg(\bigg(\frac{\varepsilon_i(\lambda)}{h}\bigg)^{2}-1\bigg)\phi
 \bigg(\frac{\varepsilon_i(\lambda)}{h}\bigg)(\dot m(W_{ib}(\lambda),\theta(\lambda,h)))^{\otimes 2}\\
 && \hskip 0.5in -\frac{1}{n}\sum\limits^{n}_{i=1}\frac{\varepsilon_i(\lambda)}{h^{3}}\phi\bigg(\frac{\varepsilon_i(\lambda)}{h}\bigg)
 \ddot m(W_{ib}(\lambda),\theta(\lambda,h))\bigg|W_{i}(\lambda)\bigg]\\
 &=& \bigg\{\frac{1}{n}\sum\limits^{n}_{i=1}\int\frac{1}{h^{2}}(t^{2}-1)\phi(t)(\dot m(W_{ib}(\lambda),\theta(\lambda,h)))^{\otimes 2}(f_{\lambda}(0|W_{i}(\lambda))+f'_{\lambda}(0|W_{i}(\lambda))ht\\
 & & +f''_{\lambda}(0|W_{i}(\lambda))\frac{h^{2}t^{2}}{2}+o_{p}(h^{2}))
 dt\\
 & &-\frac{1}{n}\sum\limits^{n}_{i=1}\int\frac{1}{h}t\phi(t)\ddot m(W_{ib}(\lambda),\theta(\lambda,h))(f_{\lambda}(0|W_{i}(\lambda))
 +f'_{\lambda}(0|W_{i}(\lambda))ht+o_{p}(h))dt\bigg\}\\
 %&=&\bigg\{\frac{2}{n}\sum\limits^{n}_{i=1}(\dot m(W_{ib}(\lambda),\theta(\lambda,h))^{\otimes %2}f''_{\lambda}(0|W_{i}(\lambda))+\frac{1}{n}\sum\limits^{n}_{i=1}\ddot m(W_{ib}
 %(\lambda),\theta(\lambda,h))f'_{\lambda}(0|W_{i}(\lambda))\bigg\}\{1+o_{p}(1)\}\\
 &=& \bigg\{\frac{1}{n}\sum\limits^{n}_{i=1}\left[(\dot m(W_{ib}(\lambda),\theta(\lambda,h)))^{\otimes 2}f''_{\lambda}(0|W_{i}(\lambda))-\ddot m(W_{ib}(\lambda),\theta(\lambda,h))
 f'_{\lambda}(0|W_{i}(\lambda))\right]\bigg\}\{1+o_{p}(1)\}\\
 &=& J(\lambda)\{1+o_{p}(1)\}
 \end{eqarray*}
by the fact $\theta(\lambda,h)\to\theta(\lambda)$, where
 \begin{eqarray*}
 J(\lambda)&=&E\bigg[ (\dot m(W_{ib}(\lambda),\theta(\lambda)))^{\otimes 2}f''_{\lambda}(0|W(\lambda))-\ddot m(W_{ib}
 (\lambda),\theta(\lambda))f'_{\lambda}(0|W(\lambda))\bigg]\\
   &=&\frac{\partial^2}{\partial\theta\partial\theta^T} E\bigg[ g(m(W(\lambda),\theta(\lambda))-m(X,\theta_0)|X)\bigg].
 \end{eqarray*}
Now let's consider the variance of each component in the matrix $\partial^{2}Q_{h}(\theta(\lambda,h))/\partial\theta\partial\theta^{T}$. For convenience, denote
  $$
    \dot m_{jk}(x,\theta)=\frac{\partial m(x,\theta)}{\partial\theta_j}\cdot\frac{\partial m(x,\theta)}{\partial\theta_k},\quad
    \ddot m_{jk}(x,\theta)=\frac{\partial^2 m(x,\theta)}{\partial\theta_j\partial\theta_k}.
  $$
For a pair $(j,k)$, $j,k=1,2,\ldots,q$,
  \begin{eqnarray*}
  &&\mathrm{Var}\bigg(\frac{\partial^{2}Q_{h}(\theta(\lambda,h))}{\partial\theta_j\partial\theta_q}\bf|W(\lambda)\bigg)\\
  &=& \mathrm{Var}\bigg[\frac{1}{n}\sum\limits^{n}_{i=1}\phi''_{h}(\vep_i(\lambda))\dot m_{jk}(W_{ib}(\lambda),\theta(\lambda, h))-\frac{1}{n}\sum\limits^{n}_{i=1}\phi'_{h}(\vep_i(\lambda))\ddot m_{jk}(W_{ib}(\lambda),\theta(\lambda, h))\bigg|
  \bW(\lambda)\bigg]\\
  &=& \frac{1}{n^{2}}\sum\limits^{n}_{i=1}\mathrm{Var}\bigg[\phi''_{h}(\vep_i(\lambda))\dot m_{jk}(W_{ib}(\lambda),\theta(\lambda, h))-
  \phi'_{h}(\vep_i(\lambda))\ddot m_{jk}(W_{ib}(\lambda),\theta(\lambda, h))\bigg|W_{i}(\lambda)\bigg]\\
  &=& \frac{1}{n^{2}}\sum\limits^{n}_{i=1}\mathrm{Var}\bigg[\frac{1}{h^{3}}\bigg(\bigg(\frac{\vep_i(\lambda)}{h}\bigg)^{2}-1\bigg)\phi\bigg(
  \frac{\vep_i(\lambda)}{h}\bigg)\dot m_{jk}(W_{ib}(\lambda),\theta(\lambda, h))\\
  &&\hskip 0.5in -\frac{\vep_i(\lambda)}{h^{3}}\phi\bigg(\frac{\vep_i(\lambda)}
  {h}\bigg) \ddot m_{jk}(W_{ib}(\lambda),\theta(\lambda, h))\bigg|W_{i}(\lambda)\bigg]\\
  &=& \frac{1}{n^{2}}\sum\limits^{n}_{i=1}\bigg\{E\bigg[\frac{1}{h^{6}}\bigg(\bigg(\frac{\vep_i(\lambda)}{h}\bigg)^{2}-1\bigg)^{2}\phi^{2}
  \bigg(\frac{\vep_i(\lambda)}{h}\bigg)\dot m_{jk}^2(W_{ib}(\lambda),\theta(\lambda, h))\\
  & &+\bigg(\bigg(\frac{\vep_i(\lambda)}{h^{3}}\bigg)^{2}\phi^{2} \bigg(\frac{\vep_i(\lambda)}{h}\bigg)\ddot m_{jk}^2(W_{ib}(\lambda),\theta(\lambda, h))\\
  & & -\frac{2}{h^{3}}\bigg(\bigg(\frac{\vep_i(\lambda)}
  {h}\bigg)^{2} -1\bigg)\phi^{2}\bigg(\frac{\vep_i(\lambda)}{h}\bigg)\bigg(\frac{\vep_i(\lambda)}{h^{3}}\bigg)\dot m_{jk}(W_{ib}
  (\lambda),\theta(\lambda, h))\ddot m_{jk}(W_{ib}(\lambda),\theta(\lambda, h))\bigg|W_i(\lambda)\bigg]\\
  &&-\bigg(E\bigg[\bigg(\frac{1}{h^{3}}\bigg(\bigg(\frac{\vep_i(\lambda)}{h}\bigg)^{2}-1\bigg)\phi\bigg(\frac{\vep_i(\lambda)}{h}\bigg) \dot m_{jk}(W_{ib}(\lambda),\theta(\lambda, h))\\
  & &
  +\frac{\vep_i(\lambda)}{h^{3}}\phi\bigg(\frac{\vep_i(\lambda)}{h}\bigg)\ddot m_{jk}(W_{ib}(\lambda),\theta(\lambda, h))\bigg|W(\lambda)\bigg]\bigg)^{2}\bigg\}\\
  &=& \bigg\{\frac{1}{nh^{5}}\frac{1}{n}\sum\limits^{n}_{i=1}\dot m_{jk}^2(W_{ib}(\lambda),\theta(\lambda, h))f_{\lambda}(0|W_{i}(\lambda))\int(t^{2}-1)^{2}\phi^{2}(t)dt\bigg\}\{1+o_{p}(1)\}\\
  &=& O_{p}\left(\frac{1}{nh^{5}}\right).
  \end{eqnarray*}
 In summary, we obtain
   \begin{eqarray*}
      \pd=O_p\left(\frac{1}{\sqrt{nh^3}}\right),\quad \pdd=J(\lambda)+o_p(1),
   \end{eqarray*}
 which imply that
   $$
     a_n\mu^T\pd=O_p(a_n^2),\quad \frac{a_n^2}{2}\mu^T\pdd\mu=O_p(a_n^2).
   $$
 Finally, note that $\|\theta^{*}(\lambda,h)-\theta(\lambda,h)\|\leq la_{n}$, then the continuity of $L_n(\theta,\mu)$ with respect to $\theta$ implies that
 \begin{eqnarray*}
   L_{n}(\theta^{*}(\lambda,h),\mu)=L_{n}(\theta(\lambda,h),\mu)+o_{p}(1).
 \end{eqnarray*}
 We can further show that
   \begin{eqarray*}
    E(L_{n}(\theta(\lambda,h),\mu)|\bW(\lambda))
    &=&\frac{1}{n}\sum\limits^{n}_{i=1}f'''_{\lambda}(0|W_i(\lambda))(\mu^T\dot m(W_{ib}(\lambda),\theta(\lambda,h)))^3\\
    & & +\frac{3}{n}\sum\limits^{n}_{i=1}f''_{\lambda}(0|W_i(\lambda))\mu^T\ddot m(W_{ib}(\lambda),\theta(\lambda,h))\mu\mu^T\dot m(W_{ib}(\lambda),\theta(\lambda,h))\\
    & & -\frac{1}{n}\sum\limits^{n}_{i=1}f'_{\lambda}(0|W_i(\lambda))F(\theta(\lambda,h),\mu)+o_p(1),
   \end{eqarray*}
 and
  \begin{eqarray*}
    \mbox{Var}(L_{n}(\theta(\lambda,h),\mu)|\bW(\lambda))= O_{p}\left(\frac{1}{nh^{7}}\right).
  \end{eqarray*}
 Therefore, we have $a_n^3L_n(\theta^*(\lambda,h),\mu)=o_p(a_n^2)$.

 Choose $\mu$ such that $\|\mu\|$ sufficiently large, then the second term in (\ref{eq5.100}) dominates other two terms. Thus, the conclusion of Theorem \ref{thm2} follows by the condition $J(\lambda)<0$.
\end{proof}

\begin{proof}[The proof of Theorem \ref{thm3}] By Taylor expansion,
 \begin{eqarray*}
    0=\frac{\partial Q_n(\hat{\theta}_{b}(\lambda))}{\partial\theta}&=&\frac{\partial Q_n(\theta(\lambda,h))}{\partial\theta}+\left[\frac{\partial^2 Q_n(\theta(\lambda,h))}{\partial\theta\partial\theta^T}+L_n\right](\hat{\theta}_{b}(\lambda)-\theta(\lambda,h)).
   \end{eqarray*}
  where
   $$
     L_n=-\frac{1}{2nh^4}\sum_{i=1}^n\left[\phi'''\left(\frac{Y_i-m(W_{ib}(\lambda),\theta^*)}{h}\right)\ddot m(W_{ib}(\lambda),\theta^*) \dot m^T(W_{ib}(\lambda),\theta^*)\right](\hat{\theta}_{b}(\lambda)-\theta(\lambda,h)),
   $$
  and
 From the proof of Theorem \ref{thm2}, we know that
  \begin{eqarray*}
    \frac{\partial^{2}Q_{h}(\theta(\lambda,h))}{\partial\theta\partial\theta^{T}}
    = J(\lambda)\{1+o_{p}(1)\}
   \end{eqarray*}
 and we can also show that $L_n=o_p(1)$. Therefore,
   \begin{eqarray}\label{eq5.5}
    0&=&\frac{1}{B}\sum_{b=1}^B\frac{\partial Q_n(\theta(\lambda,h))}{\partial\theta}+J(\lambda)\{1+o_{p}(1)\}\left(\frac{1}{B}\sum_{b=1}^B
           \hat{\theta}_{b}(\lambda)-\theta(\lambda,h)\right)\nonumber\\
      &=&\frac{1}{B}\sum_{b=1}^B\frac{\partial Q_n(\theta(\lambda,h))}{\partial\theta}+J(\lambda)\{1+o_{p}(1)\}\left(\hat{\theta}(\lambda)-\theta(\lambda,h)\right).
   \end{eqarray}
 Define
   $$
     \xi_{in}(\lambda)=\frac{h\sqrt{h}}{B\sqrt{n}}\sum_{b=1}^B\left[
    \phi'_{h}(\epsilon_{ib}(\lambda))\dot{m}(W_{ib}(\lambda),\theta(\lambda,h))\right].
   $$
 Then from (\ref{eq5.5}), we can write
  \begin{eqarray*}
     \sum\limits^{n}_{i=1}\begin{pmatrix}
                  \xi_{in}(\lambda_1)\\ \vdots \\
                  \xi_{in}(\lambda_M)
               \end{pmatrix}=
      \sqrt{nh^{3}}\begin{pmatrix}  &J(\lambda_{1})(1+o_{p}(1))&        &\\
                                    &                          & \ddots &\\
                                    &                          &        &J(\lambda_{M})(1+o_{p}(1))\\
                \end{pmatrix}
                \begin{pmatrix}
                             (\hat{\theta}(\lambda_{1})-\theta(\lambda_{1},h))\\
                                                \vdots                          \\
                             (\hat{\theta}(\lambda_{M})-\theta(\lambda_{M},h))\\
               \end{pmatrix}
     \end{eqarray*}
 Note that $(\xi_{in}^T(\lambda_1),\ldots,\xi_{in}^T(\lambda_M))^T, i=1,2,\ldots,n$ are independent and identically distributed random vectors. In the following, we shall show that the left hand side of the above equality
 are jointly asymptotically normal. By Wold technique, it is sufficient to show that for any $q$-dimensional real vectors, $\ba_1,\ldots,\ba_M$, $\sum_{i=1}^n\sum_{j=1}^M\ba_j^T\xi_{in}(\lambda_j)$ is asymptotically normal. For this purpose, we shall check the Lyapunov condition.

 By $C_r$-inequality and routing calculation, we obtain
    \begin{eqarray*}
      &&nE\left(\bigg|\sum_{j=1}^M\ba_j^T\xi_{in}(\lambda_j)\bigg|^{3}\right)\leq nM^{2}\sum\limits^{M}_{j=1}\big(E|\ba_j^T\xi_{in}(\lambda_j)|^{3}\big)\\
      &\leq& nM^{2}\left(\frac{h\sqrt{h}}{B\sqrt{n}}\right)^3\sum\limits^{M}_{j=1}\sum_{b=1}^B
      \big(E|\phi'_{h}(\epsilon_{ib}(\lambda_j))\ba_j^T\dot{m}(W_{ib}(\lambda_j),\theta(\lambda_j,h))
      |^{3}\big)\\
      %&=& \frac{M^{2}h^{9/2}}{B^2\sqrt{n}}\sum\limits^{M}_{j=1}
%      \big(E|\phi'_{h}(\epsilon_{ib}(\lambda_j))\ba_j^T\dot{m}(W_{ib}(\lambda_j),\theta(\lambda_j,h))
%      |^{3}\big)\\
     &=& \frac{M^{2}}{B^2h\sqrt{nh}}\sum\limits^{M}_{j=1} \bigg[E\bigg|\phi'\bigg(\frac{\epsilon_{ib}(\lambda_{j})}{h}\bigg)
     \ba_j^T\dot{m}(W_{ib}(\lambda_{j}),\theta(\lambda,h))\bigg|^{3}\bigg]\\
     &=& \frac{M^{2}}{B^2h\sqrt{nh}}\sum\limits^{M}_{j=1} E\int\bigg|\phi'\bigg(\frac{v}{h}\bigg)\bigg|^{3}
     |\ba_j\dot{m}(W_{ib}(\lambda_{j}),\theta(\lambda,h))|^3f_{\lambda}(v|W_{i}(\lambda))dv\\
     &=& \frac{M^{2}}{B^2\sqrt{nh}}\sum\limits^{M}_{j=1} E\int |\phi'(t)|^{3}
     |\ba_j\dot{m}(W_{ib}(\lambda_{j}),\theta(\lambda,h))|^3f_{\lambda}(th|W_{i}(\lambda))dt=O\left(\frac{1}{\sqrt{nh}}\right)
     \end{eqarray*}
 by the continuity of $\dot m$, $f_\lambda$ and convergence of $\theta(\lambda, h)$ to $\theta(\lambda)$. On the other hand, we know that
  \begin{eqarray*}
   &&\mbox{Var}(\sum_{j=1}^M\ba_j^T\xi_{in}(\lambda_j))=\sum_{j=1}^M\sum_{k=1}^M\ba_j^T\mbox{Cov}(\xi_{in}(\lambda_j),\xi_{in}(\lambda_k))\ba_k\\
   &=&\sum_{j=1}^M\ba_j^T\mbox{Cov}(\xi_{in}(\lambda_j),\xi_{in}(\lambda_j))\ba_j+
   \sum_{j\neq k=1}^M\ba_j^T\mbox{Cov}(\xi_{in}(\lambda_j),\xi_{in}(\lambda_k))\ba_k.
  \end{eqarray*}

 Note that
 \begin{eqarray*}
     &&\mbox{Cov}(\xi_{in}(\lambda_j),\xi_{in}(\lambda_j))\\
   %  &=& \mbox{Cov}\left(
   %  \frac{h\sqrt{h}}{B\sqrt{n}}\sum_{b=1}^B\left[
   % \phi'_{h}(\epsilon_{ib}(\lambda_j))\dot{m}(W_{ib}(\lambda_j),\theta(\lambda_j,h))\right],
   %  \frac{h\sqrt{h}}{B\sqrt{n}}\sum_{b=1}^B\left[
   % \phi'_{h}(\epsilon_{ib}(\lambda_j))\dot{m}(W_{ib}(\lambda_j),\theta(\lambda_j,h))\right]
   % \right)\\
    &=&\frac{h^3}{nB^2}\sum_{b=1}^B\sum_{c=1}^B
      \mbox{Cov}\left(\phi'_{h}(\epsilon_{ib}(\lambda_j))\dot{m}(W_{ib}(\lambda_j),\theta(\lambda_j,h)),
        \phi'_{h}(\epsilon_{ic}(\lambda_j))\dot{m}(W_{ic}(\lambda_j),\theta(\lambda_j,h))
      \right)\\
      &=&\frac{h^3}{nB^2}\sum_{b=1}^B
      \mbox{Cov}\left(\phi'_{h}(\epsilon_{ib}(\lambda_j))\dot{m}(W_{ib}(\lambda_j),\theta(\lambda_j,h)),
        \phi'_{h}(\epsilon_{ib}(\lambda_j))\dot{m}(W_{ib}(\lambda_j),\theta(\lambda_j,h))
      \right)\\
      &&+\frac{h^3}{nB^2}\sum_{b\neq c=1}^B
      \mbox{Cov}\left(\phi'_{h}(\epsilon_{ib}(\lambda_j))\dot{m}(W_{ib}(\lambda_j),\theta(\lambda_j,h)),
        \phi'_{h}(\epsilon_{ic}(\lambda_j))\dot{m}(W_{ic}(\lambda_j),\theta(\lambda_j,h))
      \right)\\
      &=&\frac{h^3}{nB}\mbox{Cov}\left(\phi'_{h}(\epsilon_{i1}(\lambda_j))\dot{m}(W_{i1}(\lambda_j),\theta(\lambda_j,h)),
        \phi'_{h}(\epsilon_{i1}(\lambda_j))\dot{m}(W_{i1}(\lambda_j),\theta(\lambda_j,h))
      \right)\\
      &&+\frac{h^3(B-1)}{nB}
      \mbox{Cov}\left(\phi'_{h}(\epsilon_{i1}(\lambda_j))\dot{m}(W_{i1}(\lambda_j),\theta(\lambda_j,h)),
        \phi'_{h}(\epsilon_{i2}(\lambda_j))\dot{m}(W_{i2}(\lambda_j),\theta(\lambda_j,h))
      \right)
  \end{eqarray*}
 Now we consider the first term on the right side of the above equality.
  \begin{eqarray*}
     && \mbox{Cov}\left(\phi'_{h}(\epsilon_{ib}(\lambda_j))\dot{m}(W_{ib}(\lambda_j),\theta(\lambda_j,h)),
        \phi'_{h}(\epsilon_{ib}(\lambda_j))\dot{m}(W_{ib}(\lambda_j),\theta(\lambda_j,h))\right)\\
    &=& E\left({\phi'}^{2}_{h}(\epsilon_{ib}(\lambda_j))(\dot{m}(W_{ib}(\lambda_j),\theta(\lambda_j,h)))^{\otimes 2}\right)- \left(E\phi'_{h}(\epsilon_{ib}(\lambda_j))\dot{m}(W_{ib}(\lambda_j),\theta(\lambda_j,h))
        \right)^{\otimes 2}\\
    &=&\frac{1}{h^{3}}E\bigg[\bigg(\dot m(W_{ib}(\lambda_j),\theta(\lambda_j,h))\bigg)^{\otimes 2}f_{\lambda_j}(0|W_{ib}(\lambda_j))\nu_{2}\bigg]+o\bigg(\frac{1}{h^{3}}\bigg)\\
    & & -\left[ E\dot m(W_{ib}(\lambda_j),\theta(\lambda_j,h))f_{\lambda_j}'(0|W_{ib}(\lambda_j))\right]^{\otimes 2}+O(h).
  \end{eqarray*}
 For the second term, first denote by the non-differentiable condition, the conditional density function of $\vep_i$ given $X_i, W_{i1}(\lambda_j), W_{i2}(\lambda_j))$ is the same as the as  the conditional density function of $\vep_i$ given $X_i$. Then
  \begin{eqarray*}
    && \mbox{Cov}\left(\phi'_{h}(\epsilon_{i1}(\lambda_j))\dot{m}(W_{i1}(\lambda_j),\theta(\lambda_j,h)),
        \phi'_{h}(\epsilon_{i2}(\lambda_j))\dot{m}(W_{i2}(\lambda_j),\theta(\lambda_j,h))\right)\\
   &=&E\left(\phi'_{h}(\epsilon_{i1}(\lambda_j))\phi'_{h}(\epsilon_{i2}(\lambda_j))\dot{m}(W_{i1}(\lambda_j),\theta(\lambda_j,h))
       \dot{m}^T(W_{i2}(\lambda_j),\theta(\lambda_j,h))\right)\\
   & &-\left[E\phi'_{h}(\epsilon_{i1}(\lambda_j))\dot{m}(W_{i1}(\lambda_j),\theta(\lambda_j,h))\right]
       \left[E\phi'_{h}(\epsilon_{i2}(\lambda_j))\dot{m}^T(W_{i2}(\lambda_j),\theta(\lambda_j,h))\right]\\
   &=&\frac{1}{h^6}E\bigg[\int
         \left(\vep_i+m(X_i,\theta_0)-m(W_{i1}(\lambda_j),\theta(\lambda_j,h))\right)
         \phi\left(\frac{\vep_i+m(X_i,\theta_0)-m(W_{i1}(\lambda_j),\theta(\lambda_j,h))}{h}\right)\\
   &&  \left(\vep_i+m(X_i,\theta_0)-m(W_{i2}(\lambda_j),\theta(\lambda_j,h))\right)\phi
   \left(\frac{\vep_i+m(X_i,\theta_0)-m(W_{i2}(\lambda_j),\theta(\lambda_j,h))}{h}\right)\\
   && \hskip 0.5in \dot{m}(W_{i1}(\lambda_j),\theta(\lambda_j,h))
       \dot{m}^T(W_{i2}(\lambda_j),\theta(\lambda_j,h))
        g_\vep(\vep_i|X_i)d\vep_i\bigg]\\
   & &-\left[ E\dot m(W_{ib}(\lambda_j),\theta(\lambda_j,h))f_{\lambda_j}'(0|W_{ib}(\lambda_j))\right]^{\otimes 2}+O(h)
  \end{eqarray*}
  \begin{eqarray*}
   &=&\frac{1}{h^4}E\bigg[\int
         w\phi(w) \left(wh+m(W_{i1}(\lambda_j),\theta(\lambda_j,h))-m(W_{i2}(\lambda_j),\theta(\lambda_j,h))\right)\\
   &&\hskip 0.5in
         \phi\left(w+\frac{m(W_{i1}(\lambda_j),\theta(\lambda_j,h))-m(W_{i2}(\lambda_j),\theta(\lambda_j,h))}{h}\right)\\
   && \hskip 0.5in \dot{m}(W_{i1}(\lambda_j),\theta(\lambda_j,h))
       \dot{m}^T(W_{i2}(\lambda_j),\theta(\lambda_j,h))\\
   && \hskip 0.5in     g_\vep(wh+m(X_i,\theta_0)-m(W_{i1}(\lambda_j),\theta(\lambda_j,h))|X_i)dw\bigg]\\
   & &\hskip 0.5in -\left[ E\dot m(W_{ib}(\lambda_j),\theta(\lambda_j,h))f_{\lambda_j}'(0|W_{ib}(\lambda_j))\right]^{\otimes 2}+O(h)\\
   &=&\frac{1}{h^4}E\bigg[\iiiint
         w\phi(w) \left(wh+m(X_i+u+\sqrt{\lambda_j}v_1,\theta(\lambda_j,h))-m(X_i+u+\sqrt{\lambda_j}v_2,\theta(\lambda_j,h))\right)\\
   &&\hskip 0.5in
         \phi\left(w+\frac{m(X_i+u+\sqrt{\lambda_j}v_1,\theta(\lambda_j,h))-m(X_i+u+\sqrt{\lambda_j}v_2,\theta(\lambda_j,h))}{h}\right)\\
   && \hskip 0.5in \dot{m}(X_i+u+\sqrt{\lambda_j}v_1,\theta(\lambda_j,h))
       \dot{m}^T(X_i+u+\sqrt{\lambda_j}v_2,\theta(\lambda_j,h))\\
   && \hskip 0.5in     g_\vep(wh+m(X_i,\theta_0)-m(X_i+u+\sqrt{\lambda_j}v_1,\theta(\lambda_j,h))|X_i)\phi(u)\phi(v_1)\phi(v_2)dwdudv_1dv_2\bigg]\\
   & &\hskip 0.5in -\left[ E\dot m(W_{ib}(\lambda_j),\theta(\lambda_j,h))f_{\lambda_j}'(0|W_{ib}(\lambda_j))\right]^{\otimes 2}+O(h)\\
   &=&\frac{1}{h^3}E\bigg[\iiiint
         w\phi(w)\cdot\\
   && \hskip 0.5in      \left(wh+m(X_i+u+\sqrt{\lambda_j}(v_2+vh),\theta(\lambda_j,h))-m(X_i+u+\sqrt{\lambda_j}v_2,\theta(\lambda_j,h))\right)\\
   &&\hskip 0.5in
         \phi\left(w+\frac{m(X_i+u+\sqrt{\lambda_j}(v_2+vh),\theta(\lambda_j,h))-m(X_i+u+\sqrt{\lambda_j}v_2,\theta(\lambda_j,h))}{h}\right)\\
   && \hskip 0.5in \dot{m}(X_i+u+\sqrt{\lambda_j}(v_2+vh),\theta(\lambda_j,h))
       \dot{m}^T(X_i+u+\sqrt{\lambda_j}v_2,\theta(\lambda_j,h))\\
   && \hskip 0.5in     g_\vep(wh+m(X_i,\theta_0)-m(X_i+u+\sqrt{\lambda_j}(v_2+vh),\theta(\lambda_j,h))|X_i)\\
   && \hskip 0.5in \phi(u)\phi(v_2+vh)\phi(v_2)dwdudvdv_2\bigg]\\
   && \hskip 0.5in -\left[ E\dot m(W_{ib}(\lambda_j),\theta(\lambda_j,h))f_{\lambda_j}'(0|W_{ib}(\lambda_j))\right]^{\otimes 2}+O(h)\\
   &=&\frac{1}{h^2}E\bigg[\iiiint
         w\phi(w) \sqrt{\lambda_j}v^T\left(wh+m'(X_i+u+\sqrt{\lambda_j}v_2+\tilde\lambda_j,\theta(\lambda_j,h))\right)\\
   &&\hskip 0.5in
         \phi\left(w+v^Tm'(X_i+u+\sqrt{\lambda_j}v_2+\tilde\lambda_j,\theta(\lambda_j,h))\right)\\
   && \hskip 0.5in \dot{m}(X_i+u+\sqrt{\lambda_j}(v_2+vh),\theta(\lambda_j,h))
       \dot{m}^T(X_i+u+\sqrt{\lambda_j}v_2,\theta(\lambda_j,h))\\
   && \hskip 0.5in     g_\vep(wh+m(X_i,\theta_0)-m(X_i+u+\sqrt{\lambda_j}(v_2+vh),\theta(\lambda_j,h))|X_i)
     \end{eqarray*}

  \begin{eqarray*}
   && \hskip 0.5in \phi(u)\phi(v_2+vh)\phi(v_2)dwdudvdv_2\bigg]\\
   & &\hskip 0.5in -\left[ E\dot m(W_{ib}(\lambda_j),\theta(\lambda_j,h))f_{\lambda_j}'(0|W_{ib}(\lambda_j))\right]^{\otimes 2}+O(h)\\
   &=&\frac{1}{h^2}E\bigg[\iiiint
         w\phi(w) \sqrt{\lambda_j}v^T\left(m'(X_i+u+\sqrt{\lambda_j}v_2,\theta(\lambda_j))\right)\\
   &&\hskip 0.5in
         \phi\left(w+v^Tm'(X_i+u+\sqrt{\lambda_j}v_2,\theta(\lambda_j))\right)\\
   && \hskip 0.5in \dot{m}(X_i+u+\sqrt{\lambda_j}v_2,\theta(\lambda_j))
       \dot{m}^T(X_i+u+\sqrt{\lambda_j}v_2,\theta(\lambda_j))\\
   && \hskip 0.5in     g_\vep(m(X_i,\theta_0)-m(X_i+u+\sqrt{\lambda_j}v_2,\theta(\lambda_j))|X_i)\\
   && \hskip 0.5in \phi(u)\phi(v_2)\phi(v_2)dwdudvdv_2\bigg]\\
   & &\hskip 0.5in -\left[ E\dot m(W_{ib}(\lambda_j),\theta(\lambda_j,h))f_{\lambda_j}'(0|W_{ib}(\lambda_j))\right]^{\otimes 2}+O(h)
  \end{eqarray*}
 which is the order of $o(1/h^3)$.  Therefore, we have
   \begin{equation}\label{eq5.6}
     \mbox{Cov}(\xi_{in}(\lambda_j),\xi_{in}(\lambda_j))=\frac{1}{nB}E\bigg[\bigg(\dot m(W(\lambda_j),\theta(\lambda_j,h))\bigg)^{\otimes 2}f_{\lambda}(0|W(\lambda_j))\nu_{2}\bigg]+o\bigg(\frac{1}{n}\bigg)
   \end{equation}
  uniformly for all $i$ and $\lambda_j$.

 For $j\neq k$, similar to the above derivation,  we have
   \begin{eqarray}\label{eq5.7}
     &&\mbox{Cov}(\xi_{in}(\lambda_j),\xi_{in}(\lambda_k)) \nonumber\\
      &=&\frac{h^3}{nB}\mbox{Cov}\left(\phi'_{h}(\epsilon_{i1}(\lambda_j))\dot{m}(W_{i1}(\lambda_j),\theta(\lambda_j,h)),
        \phi'_{h}(\epsilon_{i1}(\lambda_k))\dot{m}(W_{i1}(\lambda_k),\theta(\lambda_k,h))
      \right)\nonumber\\
      &&+\frac{h^3(B-1)}{nB}
      \mbox{Cov}\left(\phi'_{h}(\epsilon_{i1}(\lambda_j))\dot{m}(W_{i1}(\lambda_j),\theta(\lambda_j,h)),
        \phi'_{h}(\epsilon_{i2}(\lambda_k))\dot{m}(W_{i2}(\lambda_k),\theta(\lambda_k,h))
      \right)\nonumber\\
      &=&o(1/n)
  \end{eqarray}
 uniformly for all $i$, $\lambda_j$ and $\lambda_k$.

 From (\ref{eq5.6}) and (\ref{eq5.7}), we eventually obtain
   \begin{eqarray*}
   \mbox{Var}(\sum_{j=1}^M\ba_j^T\xi_{in}(\lambda_j))&=&\sum_{j=1}^M\ba_j^T\frac{h^3}{nB}\left[\frac{1}{h^{3}}E\bigg[\bigg(\dot m(W(\lambda_j),\theta(\lambda_j,h))\bigg)^{\otimes 2}f_{\lambda}(0|W(\lambda_j))\nu_{2}\bigg]+o\bigg(\frac{1}{h^{3}}\bigg)\right]\ba_j\\
   &&+\sum_{j\neq k=1}^M\ba_j^T\frac{h^3}{nB}o\bigg(\frac{1}{h^{3}}\bigg)\ba_k\\
   &=&\frac{1}{nB}\sum_{j=1}^M\ba_j^T\left[E\bigg[\bigg(\dot m(W(\lambda_j),\theta(\lambda_j,h))\bigg)^{\otimes 2}f_{\lambda}(0|W(\lambda_j))\nu_{2}\bigg]\right]\ba_j+o\left(\frac{1}{n}\right)
  \end{eqarray*}
 which further implies
   \begin{eqarray*}
     \mbox{Var}\left[\sum_{i=1}^n\sum_{j=1}^M\ba_j^T\xi_{in}(\lambda_j)\right]=
     \frac{1}{B}\sum_{j=1}^M\ba_j^T\left[E\bigg[\bigg(\dot m(W(\lambda_j),\theta(\lambda_j,h))\bigg)^{\otimes 2}f_{\lambda}(0|W(\lambda_j))\nu_{2}\bigg]\right]\ba_j+o(1).
   \end{eqarray*}

 By Lyapunov condition CLT, we obtain
   $$
     \sum_{i=1}^n\sum_{j=1}^M\ba_j^T\xi_{in}(\lambda_j)
     \Longrightarrow N\left(0, \frac{1}{B}\sum_{j=1}^M\ba_j^T\left[E\bigg[\bigg(\dot m(W(\lambda_j),\theta(\lambda_j,h))\bigg)^{\otimes 2}f_{\lambda}(0|W(\lambda_j))\nu_{2}\bigg]\right]\ba_j\right)
   $$
 which implies, by Wold technique,
    \begin{eqarray*}
     \sum\limits^{n}_{i=1}\left(
                  \xi_{in}^T(\lambda_1),\cdots,
                  \xi_{in}^T(\lambda_M)
               \right)^T\Longrightarrow
      N\left(0, \Pi(\Lambda)\right)
     \end{eqarray*}
 with
   $$
     \Pi(\Lambda)=\mbox{diag}\left(B^{-1}E\bigg[\bigg(\dot m(W(\lambda_j),\theta(\lambda_j,h))\bigg)^{\otimes 2}f_{\lambda}(0|W(\lambda_j))\nu_{2}\bigg]\right).
   $$
 Denote $J(\Lambda)=\mbox{diag}\left(J(\lambda_{1}),\ldots, J(\lambda_{M})\right)$. Then we eventually obtain that
 \begin{eqarray*}
    \sqrt{nh^{3}}(\hat\theta(\Lambda)-\theta(\Lambda,h))\Longrightarrow N(0, J^{-1}(\Lambda)\Pi(\Lambda)J^{-1}(\Lambda))
 \end{eqarray*}
 Combining the result from Theorem \ref{thm1}, Theorem \ref{thm3} follows.
 \end{proof}\vskip 0.2in

 {\it Acknowledgement}: Jianhong Shi's research is supported by the Open Research Fund of Key laboratory of Advanced Theory and Application in Statistics and Data Science (East China Normal University), Ministry of Education.

\bibliographystyle{elsarticle-harv}
\bibliography{zhang1}

\end{document}